\documentclass[aps,twocolumn,superscriptaddress]{revtex4-1}
\usepackage{graphics} 
\usepackage{epsfig} 
\usepackage{amsmath}
\usepackage{color}

\begin{document}
\title{Microscopic structure of electromagnetic whistler wave damping by kinetic mechanisms in hot magnetized Vlasov plasmas}
\author{Anjan Paul}
\author{Devendra Sharma}
\affiliation{Institute for Plasma Research, Bhat, Gandhinagar, India, 382428}
\affiliation{Homi Bhabha National Institute, Training School Complex, Anushaktinagar, Mumbai 400094, India}

\date{\today}

\begin{abstract}
The kinetic damping mechanism of low frequency transverse perturbations 
propagating parallel to the magnetic field in a magnetized warm electron 
plasma is simulated by means of electromagnetic (EM) Vlasov simulations. 
The short-time-scale damping of the electron magnetohydrodynamic 
whistler perturbations and underlying physics of finite electron 
temperature effect on its real frequency are recovered rather 
deterministically, and analyzed. The damping arises from an interplay between 
a global (prevailing over entire phase-space) and the more familiar 
resonant-electron-specific kinetic damping mechanisms, both of which 
preserve entropy but operate distinctly by leaving their characteristic
signatures on an initially coherent finite amplitude modification of the 
warm electron equilibrium distribution.
The net damping results from a {\em deterministic} thermalization, or {\em 
phase-mixing} process, largely supplementing the resonant acceleration of 
electrons at shorter time scales, relevant to short-lived turbulent 
EM fluctuations. A kinetic model for the evolving initial transverse EM 
perturbation is presented and applied to signatures of the whistler wave
phase-mixing process in simulations.
\end{abstract}



\maketitle
\section{Introduction \label{introduction}}
The electromagnetic turbulence prevails in
collective excitations 
of charged matter interacting with, and by means of, the electromagnetic field
over a vast range of spatiotemporal scales, usually terminated by dissipation at the finer 
scales.
The solar-wind spectrum, for example, shows that beyond a frequency 
{\em breakpoint} a deviation exists from the inertial rage characterized by 
exponent -5/3 of power law \cite{Borovsky_2012,Podesta_2007}. In one of the 
plousible scenarios, the whistler 
fluctuations can be the fundamental mode 
and central means of dissipation in this weak turbulence regime \cite{Gary_2010}. 
A steepening present in the spectrum 
leading to considerably high spectral exponents ($\sim$2-3)
suggests presence of considerable damping alongside to the intra-spectral 
energy transfer \cite{Gary_2010}.
Besides by conversion to electrostatic modes \cite{xu:x}, 
damping by kinetic transverse wave-particle interaction must 
operate on the short lived excitations 
\cite{helliwell:ra,xiao:f,thorne:rm,chen2013improved} initiated by sponteneous 
field fluctuations. 
Fresh perturbations, so triggered, 
excite warm plasma eigenmodes by leaving long lasting 
remnants of their initially enforced 
phase-space 
structure 
in the memory of nonthermal kinetic distributions \cite{krall:na}. The asymptotic long-time 
solutions of the collisionless kinetic formulation 
\cite{landau:ld,landau:ld1946} applied to them 
therefore have large 
scope of sophistication by admitting a strong {\em deterministic thermalization}, or phase-mixing \cite{krall:na}, alongside the damping evaluated in usual time asymptotic, 
$t\rightarrow 0$ limit.

The general kinetic evolution produced collisionless damping of 
electromagnetic fluctuations \cite{lutomirski1970physical} involves a rather 
complicated phase-space dynamics and
is most accessible by
deterministic Vlasov simulations \cite{fijalkow1999numerical,mangeney:a}. 
Only a limited number of studies 
have rather deterministically simulated the dynamics of the transverse
electromagnetic excitations and their damping/stability in a hot collisionless 
magnetized plasma \cite{Palodhi_2010,valentini:f}, even as the process 
remains critical for determining 
the operational state of turbulence and the transport associated with it 
both in space plasmas \cite{Gary_2010,shuster:jr} as well as in modern magnetic confinement 
fusion experiments \cite{fulop2009magnetic}.

In the collisionless limit, the modifications made to temperature, or width,
of an initially equilibrium warm electron velocity distribution produce a 
higher order correction to the resonant wave particle interaction process.
The analytical model predicts a related downward shift in the whistler wave 
frequency in collisionless plasmas with hotter electrons 
\cite{chen2013improved,schreiner:c,schreiner2016numerical}. The recovered 
strength of damping due to wave particle interaction however remains 
underestimated in comparison with that produced by the computer simulations 
implemented with reasonably low collisionality. 

This paper addresses above aspects of kinetic whistler damping mechanism,
subsequent to the recovery of general 
electromagnetic modes of a magnetized plasma and their dispersive 
characterization in our flux-balance based 
\cite{mandal:dconf,fijalkow1999numerical} Vlasov 
simulations.  
This is accompanied by 
illustration of its detailed {\em phase-spatiotemporal} evolution. 
The interaction of electromagnetic modes, propagating parallel to the magnetic 
field with the resonant particles is studied, particularly recovering the 
damping of the whistler waves via full kinetic mechanism and comparison of 
the simulation results with those analytically prescribed in the linear 
Landau theory limit \cite{landau:ld,krall:na}. 
Presented simulations and analysis enter the finer regime of kinetic 
phase-mixing of the electromagnetic mode uniquely achievable by Vlasov 
simulations.
We qualitatively recover the phase-mixing effects showing the dominance 
of frequency ${\bf v\cdot k}$ of the ballistic term ($\propto \exp(ikvt)$) 
\cite{krall:na} accounting for short time response, in addition to 
time asymptotic 
Landau damping results that are routinely applied to turbulent electromagnetic 
fluctuations of sufficiently short life time.
First quantitative analysis of the {\em phase-spatiotemporal} evolution of a 
4D electron phase-space distribution perturbation associated with the 
transverse electromagnetic whistler mode simulated in a hot magnetized 
Vlasov plasma is presented.

The present paper is organized as follows. In Sec.~\ref{model} the electromagnetic Vlasov-Maxwell model considered for the simulation is presented. Vlasov simulation implementation and its characterization by dispersion of transverse EM perturbation is presented in Sec.~\ref{simulations}. In Sec.~\ref{whistler-damping} the simulations of Landau damping of transverse whistler wave perturbations in a warm electron plasma is presented followed by their comparison with analytical results. Based on the evolution of simulated perturbations, the signatures of phase-mixing supplemented landau damping of the initial perturbation are presented in Sec.~\ref{phase-space}. The kinetic model for initial transverse EM perturbations is presented and solved in Sec.~\ref{initial-value-problem}. Conclusions are presented in Sec.~\ref{conclusion}.
\section{The electromagnetic Vlasov plasma model \label{model}}
The electromagnetic magnetized plasma modes simulated in this paper 
follow pure kinetic formulation and are well represented by the solutions of 
collisionless, fully nonlinear Vlasov equation for species $\alpha$,
\begin{eqnarray}
	\frac{\partial f_{\alpha}}{\partial t} + {\bf v}\cdot\frac{\partial f_{\alpha}}{\partial {\bf x}}+
	\frac{q_{\alpha}}{m_{\alpha}}\left({\bf E}+\frac{{\bf v}\times {\bf B}}{c}\right)
	\frac{\partial f_{\alpha}}{\partial {\bf v}}=0.
\label{Vlasov_eq}
\end {eqnarray}
The evolution of electric field ${\bf E}$ and magnetic field 
${\bf B}$ for the electromagnetic processes follows the Maxwell's equations,
\begin{eqnarray}\label{M1}
		\nabla\times{\bf B}&=&\frac{1}{c}\frac{\partial {\bf E}}{\partial t}+\frac{4\pi}{c}{\bf J},\\\label{M2}
		\nabla\times{\bf E}&=&-\frac{1}{c}\frac{\partial {\bf B}}{\partial t},\\\label{M3}
	\nabla\cdot{\bf B}&=&0
\end{eqnarray}
	where the current density ${\bf J}$ is described by 
\begin{eqnarray}
	{\bf J}=\sum_{\alpha}\int_{-\infty}^{\infty} d{\bf v}~q_{\alpha}{\bf v}f_{\alpha}
	\label{def-j}
\end{eqnarray}
Consistent with the electron magnetohydrodynamic regime excitations, we use 
an externally applied constant magnetic field ${\bf B}_{0}$ and infinitely 
massive ions. The subscript for species $\alpha$ therefore only assumes 
value, $e$, representing electrons (and henceforth omitted), 
which are the only mobile species and contributing to the perturbation.

The full nonlinear kinetic model (\ref{Vlasov_eq})-(\ref{def-j}) is 
implemented for the case of
waves propagating parallel to an applied magnetic field (${\bf B_{0}\|k}$) 
where principle modes 
are high and low frequency right and left handed circularly polarized modes.
This set up includes the whistler waves that are right handed polarized and 
propagate in the frequency range $\Omega_{ci}<\omega<\Omega_{ce}$, where 
$\Omega_{c\alpha}$ is the gyrofrequency of the species $\alpha$.
\section{Kinetic simulations of electromagnetic waves in magnetized 
	warm Vlasov plasma\label{simulations}}
The simulations presented in this analysis progress by numerically evolving 
the magnetized plasma (electron) velocity distribution function $f$ 
according to the dynamics of the phase-fluid flow \cite{krall:na} which is 
governed by the collisionless 
Vlasov equation (\ref{Vlasov_eq}), and associated Maxwell's equations 
(\ref{M1})-(\ref{M3}).
The 4-dimensional (4D $\equiv$ 1x-3v) phase-space simulations are performed 
using an advanced flux balance technique 
	\cite{mandal:d2020,mandal:dconf,fijalkow1999numerical} generalized 
to simulate the electromagnetic plasma modes in a large 
range of magnetization of plasma species. 
The results of simulations are characterized first against the 
analytical dispersion relation of left and right handed circularly polarized 
low and high frequency modes and then against both analytic and 
numerical evaluation of the linear Landau damping descriptions \cite{krall:na,gary1993theory,chen2013improved}. 
\subsection{The simulation set-up}
In order to simulate the waves propagating parallel to an ambient magnetic 
field ${\bf B}_{0}$ in a wide range of frequency $\omega$ and wave vector 
${\bf k}$, we have assumed a setup 
where both the ${\bf B}_{0}$ and ${\bf k}$ are aligned to $z$-axis and the 
periodic boundary condition is used at both the boundaries of the 
one-dimensional simulation zone located between 
$z=0$ and $L$. The setup therefore assumes symmetry along both $\hat{x}$ and 
$\hat{y}$ directions with finite spread of electron velocities along these 
dimensions, besides the direction $\hat{z}$.
The three dimensional equilibrium velocity distribution for the electron 
species is considered to be a Maxwellian, 
\begin{eqnarray}
	f(z,{\bf v})=\left(\frac{1}{2\pi v_{th}^{2}}\right)^{3/2}
	\prod_{j=x,y,z}\exp{\left\{-\frac{
		(v_{j}-\langle v_{j}\rangle)^{2}
	}{2v_{th}^{2}}\right\}},
	\label{f-eq}
\end{eqnarray}
where $v_{th}=(T_{e}/m_{e})^{1/2}$ is electron thermal velocity, $T_{e}$ is 
electron temperature in energy units and $m_{e}$ is electron mass.

At time $t=0$, the equilibrium distribution is perturbed with sinusoidal 
perturbations (having variation along $\hat{z}$) 
with initial amplitudes of
$\langle{\bf v}\rangle_{1}$, ${\bf E}_{1}$ and ${\bf B}_{1}$ 
($\propto exp{(ikz)}$) consistent with 
transverse electromagnetic (right or left handed circularly polarized) 
linear magnetized plasma modes propagating with a desired $k$ 
\cite{chen:ff,helliwell:ra}. 
\subsection{Dispersion characterization of electromagnetic plasma modes \label{characterization}}
We present the dispersion relations recovered for the case of very small 
electron thermal velocity ($v_{th}=0.001$ c) simulated and its comparison 
with the corresponding analytical cold plasma dispersion relation 
\cite{stix1992waves,chen:ff}.
\begin{figure}
	\includegraphics[width=2.2 in]{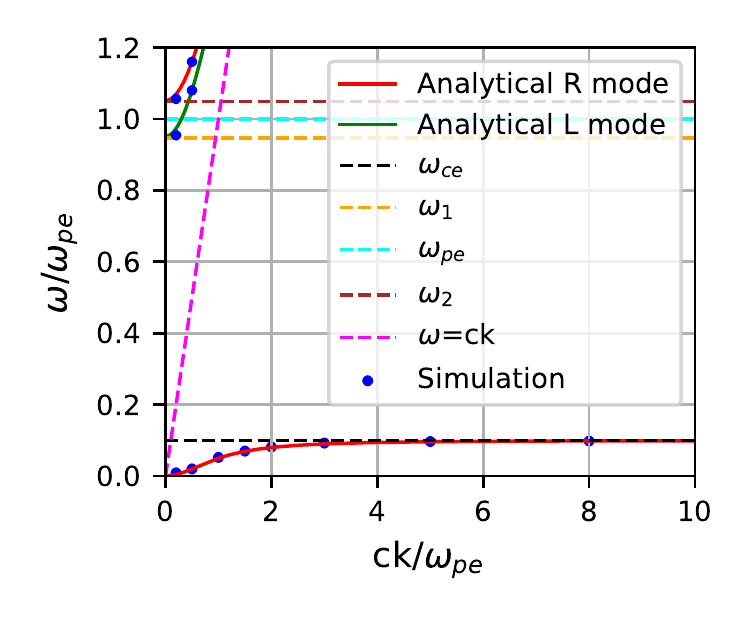}
\caption{ 
Dispersion of simulated frequency (dots) compared to the Right and Left handed 
polarized branches of the analytical dispersion relation (solid line). 
\label{dispersion}
}
\end{figure}
Specific to simulation cases presented in this section, 
we have used a 3-dimensional velocity space grid of rather moderate size
having $32\times 32\times 32$ grid points, in combination with spatial 
grid size also of $32$ grid points. 
In Fig.~\ref{dispersion}, the comparison is presented of the simulated 
values of frequency $\omega$ of the perturbation plotted as function of 
$k$, with the right and left handed circularly polarized (RHCP and LHCP) 
branches of the analytical dispersion relation in the limit of infinitely 
massive ions 
($\omega_{\rm pe}\gg\omega_{\rm pi}$),
\begin{eqnarray}
	k^{2}_{R,L}=\frac{\omega^{2}}{c^{2}}\left[
		1-\frac{\omega_{pe}^{2}}{\omega(\omega \mp \Omega_{ce})}
	\right].
	\label{dispersion-analytic}
\end{eqnarray}
Considering the parameters,
$v_{th}=0.001$ c,
the ratio of electron cyclotron 
frequency and plasma frequency $\Omega_{ce}/\omega_{pe}=0.1$ and sufficiently
small initial perturbation exclusively in electron average velocity amplitude 
$\Delta V=\langle v \rangle_{\rm 1max}-\langle v \rangle_{\rm 1min}=2\times 10^{-3}c$,
the high frequency ($\omega>\Omega_{ce}$) RHCP and LHCP mode phase velocities 
are sufficiently high for these waves to stay out of resonance with the
cold electrons chosen for this case.
Moreover, for the low frequency whistler waves excited
(dots on the red curve in the region $0<\omega<\Omega_{ce}$ in 
Fig.~\ref{dispersion}), 
the electron thermal velocity $v_{th}=0.001 c$ chosen is still sufficiently 
low for no significant resonant electron population to be available at 
the resonant velocity, $v_{z}=v_{\rm res}=(\omega-\Omega_{ce})/k$. 
In rest of this analysis we exclusively characterize this low frequency 
whistler branch of the perturbation for relatively warmer electrons 
in order to analyze its resonant damping. 
\section{Landau damping of the transverse Whistler mode \label{whistler-damping}}
\begin{figure}
	\includegraphics[width=2.2 in]{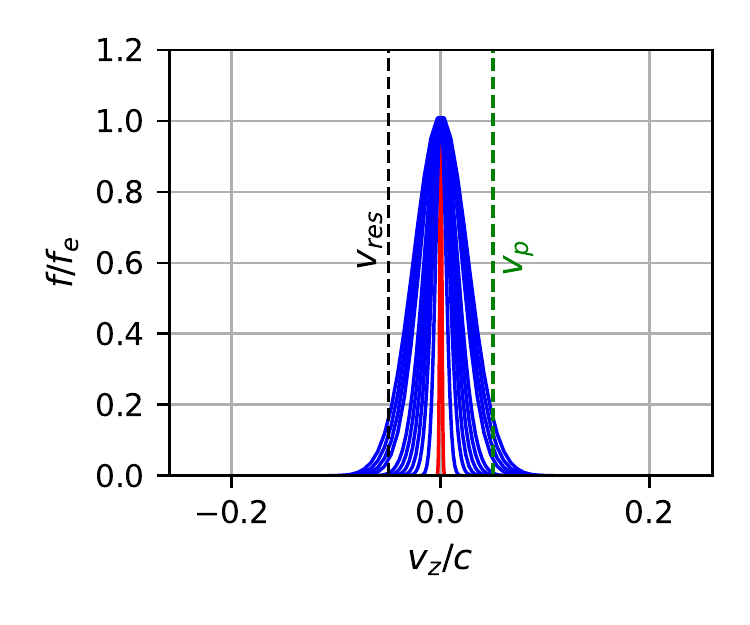}
\caption{ 
	The electron distribution function (normalized to its maximum value) 
	plotted as function of parallel velocity $
	v_{\|}\equiv v_{z}$ at $v_{x}=v_{y}=0$. The thermal 
	velocity of electrons for the curves with increasing width ranges 
	from $v_{th}=0.001 c$ (inner most profile) to $0.026 c$ (outer most 
	profile) , respectively. The vertical dotted lines indicate resonant 
	velocity $v_{\rm res}$ and phase velocity $v_{p}$ of the low 
	frequency whistler mode at $k=1.0 \omega_{p}/c$.
\label{fe}
}
\end{figure}

An advanced simulation set up, with grid size of 
$64\times 64\times 64 \times 64$ 
is implemented in the following sets of simulations, by adopting the 
electron velocity range [-0.26, 0.26] c. 
The velocity distribution of electrons as function of the parallel velocity 
$v_{\|}\equiv v_{z}$ is plotted in Fig.~\ref{fe} at $v_{x}=v_{y}=0$ 
for a range of electron thermal velocity $v_{th}=0.001 c$ (inner most profile) 
to $0.026 c$ (outer most  profile) explored in the cases presented in this 
section. The resonant velocity $v_{\rm res}=(\omega-\Omega_{ce})/k$ and 
phase velocity $v_{p}$ of the low frequency whistler mode for parameters
$k=1.0 \omega_{p}/c$ and $\Omega_{ce}/\omega_{p}=0.1$, are indicated by vertical 
dotted line. The profiles for smaller values of $v_{th}$ show the 
population of resonant electrons drops to 
negligible values such that no resonant damping is present for these cases.

\begin{figure}
	\includegraphics[width=2.0 in]{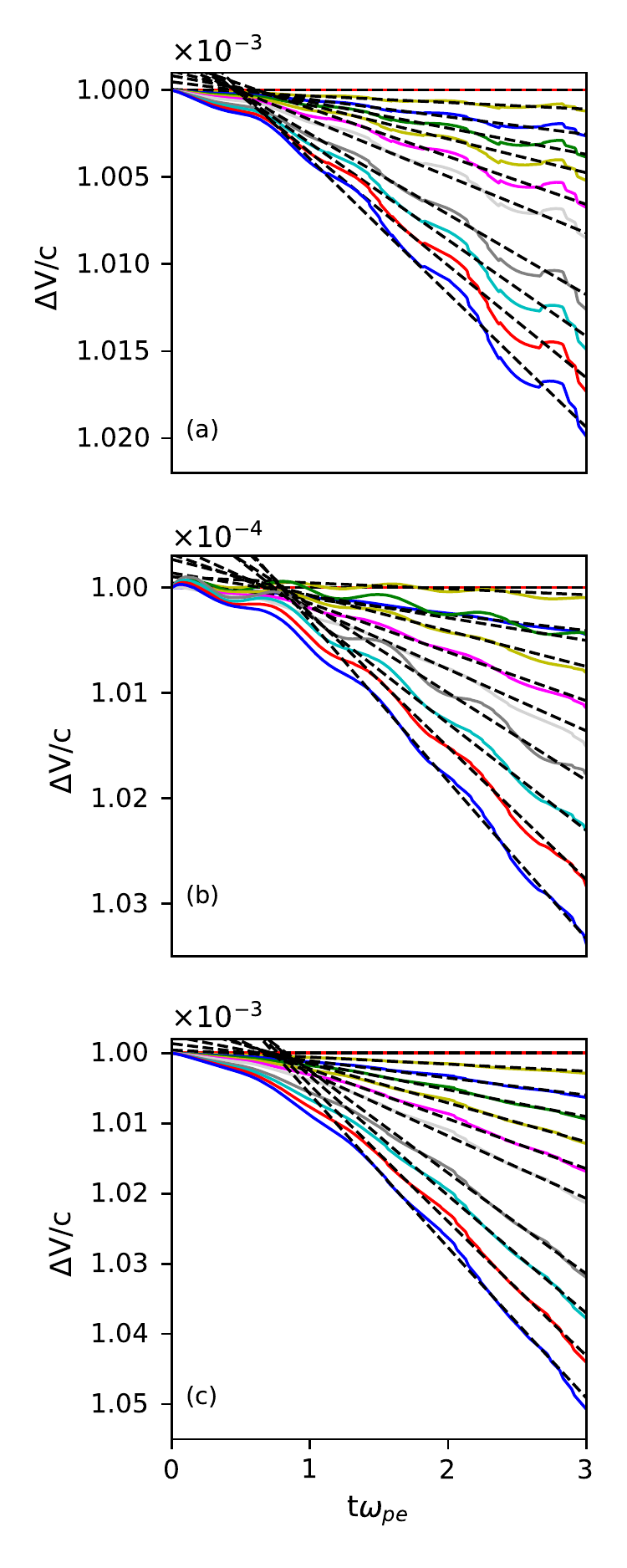}
\caption{ 
	The decay in the velocity perturbation $\Delta v$ for 
	$v_{th}=0.001 c$ (lowest damping) to $0.026 c$ (highest damping) for
	(a) $k=0.8\omega_{pe}/c$,
	(b) $k=0.897\omega_{pe}/c$ and 
	(c) $k=1.0\omega_{pe}/c$. 
	The modulation seen in some cases are residual
       RHCP and LHCP excitations
	($\omega_{R,L}>\Omega_{ce}$ satisfying (\ref{dispersion-rhcp-cold})).
\label{decay1}
}
\end{figure}

In 
Fig.~\ref{decay1}(a), 
the time evolution of the amplitude of the velocity 
perturbation $\Delta V$ is plotted in the simulations done for the range 
of electron thermal velocity  $v_{th}=0.001 c$ to $0.026 c$ for the 
value of $k=0.8\omega_{pe}/c$. Two additional set of simulations done for 
the values 
of $k=0.897$ and $k=1.0\omega_{pe}/c$ are presented in 
Fig.~\ref{decay1}(b) and 
Fig.~\ref{decay1}(c), respectively.
The time evolution of $\Delta V$ in all the cases above cover the initial 
evolution of the wave amplitude only for the time duration $\Delta 
t\sim 3\omega_{p}^{-1}$ (i.e., about a fraction of one complete cycle of the 
whistler wave with frequency $\omega<\Omega_{ce}$ while 
$\Omega_{ce}/\omega_{p}=0.1$) which is the time duration in which the linear
Landau damping rate remains reasonable estimate for the wave damping.
%
%
The short time evolution in Fig.~\ref{decay1} has sufficiently low numerical
widening of $f$ for resolving the effect of $T_{e}$ variation, which is 
varied with relatively much larger increments in this study. 
\subsection{Comparison of simulations with analytic whistler damping rates}
The damping rates of whistler velocity perturbation amplitude can be 
compared with the available analytic approximations of the Landau damping 
rate of whistlers. For the purpose of comparison we have used the analytic 
results from kinetic formulation in certain limiting cases.
The numerical evaluation of analytical expression is done in the limits 
where such approximations become unavailable. 
We begin by using the kinetic dispersion relation for the electromagnetic 
waves (\cite{bittencourt2013fundamentals} and Sec.~\ref{initial-value-problem}) given by,
\begin{eqnarray}
c^{2}k^{2}-\omega^{2} = \frac{8\pi^{2}q_{e}^{2}}{m}
\int\limits_{-\infty}^{\infty}\int\limits_{0}^{\infty}
	&&\frac{
\left[(\omega-kv_{z})\frac{\partial f_{e0}}{\partial v_{\perp}}+
kv_{\perp}
\frac{\partial f_{e0}}{\partial v_{z}}
\right]
}{(\omega-kv_{z}\pm \Omega_{ce})}\nonumber\\
	&\times& v_{\perp}^{2}dv_{\perp}dv_{z}.
	\label{kinetic-dispersion}
\end{eqnarray}
Assuming isotropy in the velocity space,
\begin{eqnarray}
	\frac{\partial f_{0}}{\partial v_{\|}}
	=\frac{\partial f_{0}}{\partial v_{\perp}},
\end{eqnarray}
additionally considering the whistler wave frequency limit 
$\Omega_{ci}\ll\omega\ll\Omega_{ce}$ and for wavelengths larger than 
electron gyroradius, $\lambda>\rho_{ce}$, we recover \cite{krall:na},
\begin{eqnarray}
	\gamma\approx -\frac{\omega_{pe}^{2}}{|k_{\|}|v_{th}}
	\frac{1}{1+k^{2}c^{2}/\omega_{r}^{2}}\exp\left(
	-\frac{\Omega_{ce}^{2}}{k_{\|}^{2}v_{th}^{2}}
	\right).
	\label{def-gamma}
\end{eqnarray}
However, for the cases explored in the present simulations have 
$\omega\lesssim \Omega_{ce}$ and
the expression (\ref{def-gamma}) underestimates the damping rate $\gamma$, 
yielding negligible values in comparison to what recovered in the simulations.
For example, for $k=1.0\omega_{pe}/c$, $v_{th}=0.026 c$, 
$\Omega_{ce}=0.1 \omega_{pe}$, $\gamma\sim 10^{-8} \omega_{pe}$, although a considerably 
higher damping ($\gamma\sim 10^{-3} \omega_{pe}$) is recovered in the 
simulations.
An estimate from (\ref{kinetic-dispersion}) in the limit 
$\omega\lesssim \Omega_{ce}$ more relevant to simulations is prescribed by 
Gary \cite{gary1993theory} in the form of a general expression for $\gamma$,
\begin{eqnarray}
	\gamma\approx -\frac{\pi}{2 k}\sum_{\alpha}\frac{\omega_{p\alpha}^{2}}{\sqrt{2\pi} v_{th \alpha}}
	\exp\left(- 
	\frac{(\omega\pm \Omega_{c\alpha})^{2}}{2 k_{\|}^{2}v_{th \alpha}^{2}}
	\right)
	\label{def-gamma2}
\end{eqnarray}
which can be used as an alternate analytic estimate to compare with theory 
the whistler damping rate recovered in the simulations.
For an even improved analytic estimate of damping rates we have also 
used the exact expression (\ref{kinetic-dispersion}) and evaluated $\gamma$ 
numerically, by the following procedure.

As in the simulation where ions are infinitely massive, 
considering electron equilibrium distribution to be Maxwellian,
\begin{eqnarray}
	f_{0} =  \left(\frac{1}{2\pi}\right)^{\frac{3}{2}} \frac{1}{v_{th \parallel e} v_{th\perp e}^{2}}\exp\left(-
	{\frac{v_{\perp}^{2}}{2 v_{th_{\perp e}}^{2}}-\frac{v_{\parallel}^{2}}{2 v_{th_{\parallel e}}^{2}}}\right),\nonumber\\
\end{eqnarray}
and substituting it in the dispersion (\ref{kinetic-dispersion}), 
one obtains \cite{chen2013improved},
\begin{equation}
\omega^{2}+\omega_{pe}^{2}I-c^{2}k^{2}=0,
	\label{dispersion1}
\end{equation}
where,
\begin{eqnarray}
%
I =\frac{v_{th\perp e}^{2}}{v_{th\parallel e}^{2}}
-1+\left[\frac{v_{th\perp e}^{2}}{v_{th\parallel e}^{2}} \frac{\omega \pm \Omega_{ce}}{\mp\Omega_{ce}}+1\right]
\frac{\mp\Omega_{ce}}{\sqrt{2}kv_{th \parallel e}}Z(\zeta),\nonumber\\
	\label{def-I}
\end{eqnarray}
and $ Z(\zeta)$ is the plasma dispersion function \cite{abramowitz+stegun}, 
\begin{equation}
  Z(\zeta)  = \frac{1}{\sqrt \pi}\int \limits_{-\infty}^{\infty} \frac{e^{-t^{2}}}{t-\zeta}dt,
\end{equation}
with the argument
\begin{equation}
\zeta=\frac{\omega \mp\Omega_{ce}}{\sqrt{2}kv_{th \parallel e}}.
\end{equation}
The upper and lower sign in Eq.~(\ref{def-I}) are for RHCP and LHCP waves, 
respectively. 
Applying dispersion (\ref{dispersion1}) to whistlers which are right 
handed polarized, and considering isotropy  of the distribution 
($v_{th \parallel}=v_{th \perp}=v_{th}$) we get (\ref{dispersion1}) in the 
form,
\begin{equation}
\omega^{2}+\omega 
	\frac{\omega_{pe}^{2}}{\sqrt 2 kv_{th}} Z(\zeta) -c^{2}k^{2}=0.
	\label{dispersion-rhcp}
\end{equation}
Note that under cold plasma approximation where imaginary part of the function 
$Z(\zeta)=i\sqrt \pi e^{-\zeta ^{2}} -\frac{1}{\zeta}$ is negligible and
$Z(\zeta)\sim-1/\zeta$, such that with 
$\zeta=(\omega-\Omega_{ce})/\sqrt 2 k v_{th}$
one readily recovers from (\ref{dispersion-rhcp}), the cold plasma dispersion 
relation for the RHCP waves,
\begin{equation}
\omega^{2}-\omega \frac{\omega_{pe}^{2}}{(\omega-\Omega_{ce})}-c^{2}k^{2}=0,
	\label{dispersion-rhcp-cold}
\end{equation}
which is the whistler dispersion relation in the limit 
$\Omega_{ci}<\omega<\Omega_{ce}$.
Considering $\omega=\omega_{r}+i\omega_{i}$, in the warm plasma RHCP 
dispersion (\ref{dispersion-rhcp}) we obtain for 
$\gamma=\omega_{i}\ll\omega_{r}$ \cite{chen2013improved,schreiner2016numerical}, 
\begin{equation}
\gamma=-\frac{\omega_{r}\sqrt \pi e^{-\zeta^{2}}\sqrt 2 k v_{th}}{[\omega_{r}Z'(\zeta)+
\sqrt 2 k v_{th}Z(\zeta)+2\omega_{r}k^{2}v_{th}^{2}]},
	\label{def-gamma1}
\end{equation}
where $Z'({\zeta})$ is the derivative of $Z({\zeta})$ with respect to $\zeta$.

\begin{figure}
	\includegraphics[width=2.3in]{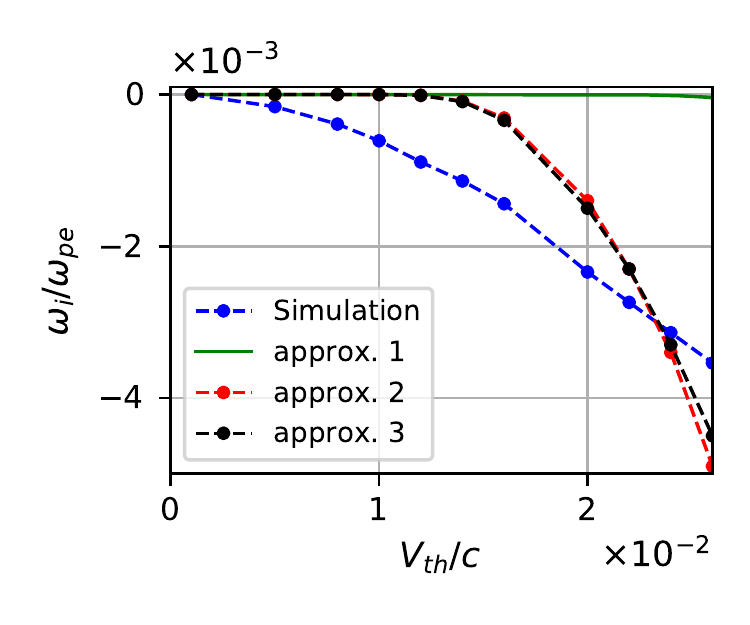}
	\caption{Comparison of simulated damping rate $\gamma$ with that obtained from analytical approximations, approx-1 (\ref{def-gamma}), approx-2 (\ref{def-gamma1}) and approx-3 (\ref{def-gamma2}) done for $k=1.0\omega_{p}/c$ or simulation box length $L=2\pi c/\omega_{p}$.
\label{comparison}
}
\end{figure}

Comparison of the damping rate recovered in simulations with the analytical 
approximations (\ref{def-gamma}), (\ref{def-gamma2}) and (\ref{def-gamma1}) 
is presented in Fig.~\ref{comparison} for the above range of $v_{th}$ values
for which simulations are performed using $k=0.897 \omega_{p}/c$ or simulation
box length $L=7 c/\omega_{p}$.
Note that for the nearly cold electron case, 
$v_{th}\rightarrow 0$, the simulations (blue dots) duly recover
a whistler propagation free of any damping (or growth), confirming that 
the deviation from this undamped propagation recovered at large electron 
temperature values (larger $v_{th}>0$) represents pure 
kinetic effects in the simulations. The simulations however show increasing 
damping from finite
$v_{th}$ values while all the other analytical approximations prescribe 
comparable damping only beyond $v_{th}>1.0\times 10^{-2}c$, as discussed 
further below.
\begin{figure}
	\includegraphics[width=2.5 in]{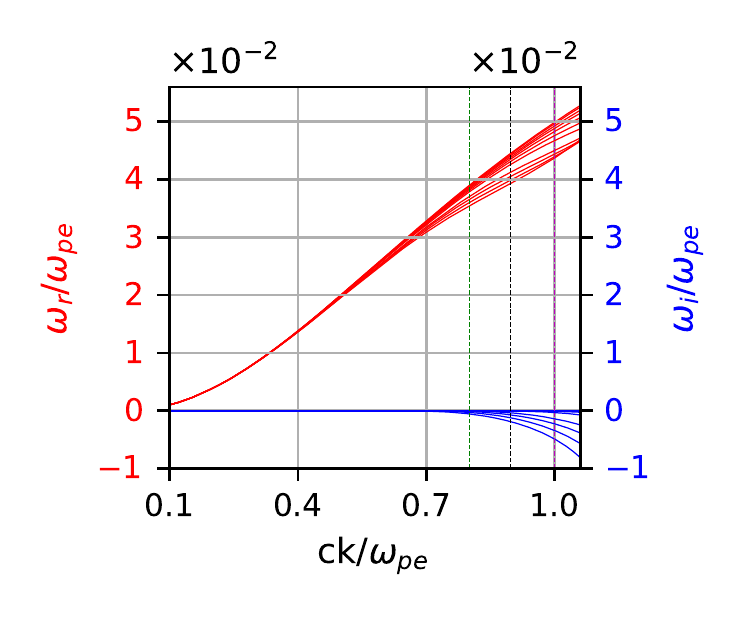}
\caption{Simulated frequency and damping rate $\gamma$ as function of 
$k$ and a range in electron thermal velocity values $v_{th}=0.001$ to $0.026$.
	The vertical dashed lines indicate $k$ values for which sets of simulations are performed by varying $v_{th}$.
\label{frequency-damping}
}
\end{figure}

Suitability of approximation (\ref{def-gamma}) for the present range is 
ruled out by the negligible resonant damping prescribed by (\ref{def-gamma}) 
as plotted in Fig.~\ref{comparison} using green line in comparison to
other estimates of $\gamma$, namely, (\ref{def-gamma2}) and (\ref{def-gamma1}) 
plotted with red and black lines, respectively. 
\begin{figure}
	\includegraphics[width=2.5 in]{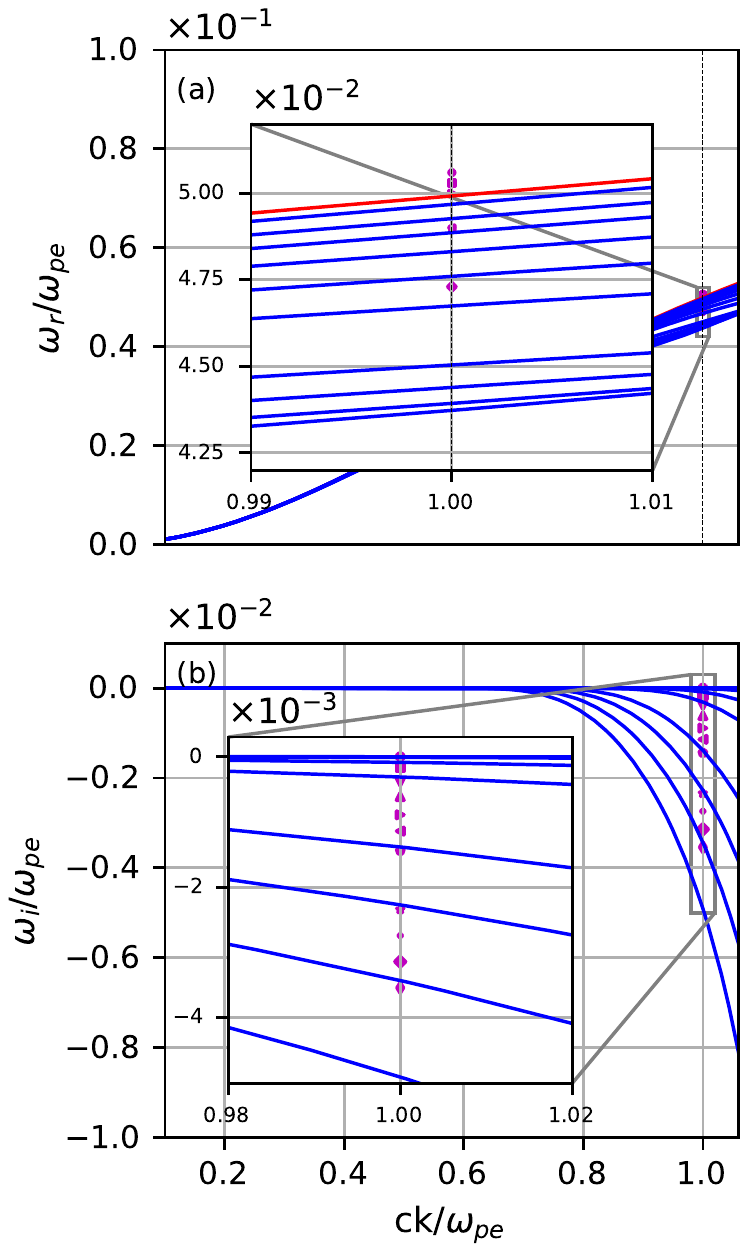}
	\caption{Comparison of simulated frequencies (a) $\omega_{r}$ and (b) $\omega_{r}$ with those obtained from analytical approximation (\ref{def-gamma1}) and corresponding $\omega_{r}$ as functions of $k$ and $v_{th}$. Markers at various $k$ values represent simulation data such that the markers and curves from top to bottom correspond to $v_{th}=0.001$ to $0.026$.  \label{comparison1} }
\end{figure}
The analytical approximation (\ref{def-gamma2}) very closely estimates 
the numerically evaluated values of the exact expression (\ref{def-gamma1}).
We note that despite no clear approximations used in them
(at least, in the numerically computed $\gamma$ from (\ref{def-gamma1}))
besides their linear origin, they still show
finite discrepancy from the simulated damping rates.
In general, the simulations recover stronger damping 
rates for the whistlers
than what prescribed by the analytical damping 
rate approximations duly obtained from the kinetic dispersion relation. 
We explore the origin of this discrepancy for whistlers excitations 
in Sec.~\ref{phase-space} based on the deterministic evolution of the 
electron distribution function in essential 4D phase-space set up available 
from the simulations. The missing physics is located in 
the small 
time evolution of the initial perturbations as asymptotic 
contributions are duly accounting for in estimating the kinetic Landau 
damping rate.
For turbulent situations, where the excited fluctuations are short lived,
the short time contributions need to be accounted for. Moreover, they
might dominate the asymptotic mechanisms in certain relevant limits.

In Fig.~\ref{frequency-damping} the analytically obtained whistler 
frequency $\omega$ and more accurate numerically computed damping rate 
from (\ref{def-gamma1}) are plotted as 
function of $k$ for several ($>10$) different values of electron thermal 
velocity in range $v_{th}=0.001$ to $0.026$. 
In Fig~\ref{comparison1}, the $\omega_{r}$ and 
$\gamma$ values are separately compared with analytical values where 
the markers represent simulated data and variation is done in both $k$ 
and $v_{th}$ values. 
As in Fig~\ref{comparison1}, simulations confirm the analytical prediction 
that the $\omega_{r}$ slightly reduces at larger electron temperature, though 
the analytical $\omega_{r}$ overestimates this reduction for larger $v_{th}$. 
The comparison of $\gamma$ presented in Fig.~\ref{comparison1} shows that 
while at $k=0.8 \omega_{p}/c$ the damping rate in simulation is recovered to 
be stronger than the corresponding analytic values, at larger $k$ values, 
$0.897~\omega_{p}/c$ and $1~\omega_{p}/c$, the overlap in the analytic and 
simulated $\gamma$ values is better, 
especially at larger $v_{th}$ values. It can also be noted that at 
smaller $k$, analytical expressions underestimate of the damping and the
spread in kinetic $\gamma$ values is larger than the analytical value spread.
At larger $k$, on the other hand, the analytical approximation overestimates 
$\gamma$ and the kinetically obtained $\gamma$ values have a narrower spread.
%
\section{Phase-mixing supplemented damping of the EMHD perturbations \label{phase-space}}
Relatively larger damping of the whistler perturbations at smaller $k$ in 
comparison to the analytical estimates (\ref{def-gamma}), (\ref{def-gamma2}) 
and (\ref{def-gamma1}) is interpreted based on the deterministic evolution of 
the electron distribution function governed by the Vlasov equation as 
simulated and analyzed in this section.
In warm electron plasmas the initial damping is found to have considerable 
contribution from the corresponding deterministic phase-mixing process of the 
fresh (initial) perturbations made to the plasma, confirming to a propagating 
whistler wave. 
Note that the term {\em phase-mixing} used here refers to the varied 
response of 
the electrons of different velocities (from a wider spread about mean) 
to the propagating wave perturbation. 
The deterministic deformations of the velocity distribution function that 
evolve antisymmetrically about a reference velocity (distinct for 
electrostatic and electromagnetic waves) in the phase-space however 
introduce incoherence between the substructures of
observable macroscopic 
spatio-temporal variations, or collective wave perturbations. 
The resultant weakening of the perturbations 
translates in to an
additional damping (than what caused by the resonant electrons) 
detectable in the deterministic Vlasov simulation data.
At shorter time scales this contribution supplements, and modifies, 
the damping of the (transverse) wave, producing enhanced
damping rates in comparison to the pure resonant damping 
essentially estimated considering the long time limit \cite{krall:na}.
Clear signatures of both resonant and non resonant processes are visible 
in the simulated 3-dimensional electron velocity distribution function 
evolving in the phase-space.
\subsection{Phase-space evolution of whistler wave perturbation in the Vlasov simulations \label{mixing}}
\begin{figure}
	\includegraphics[width=3.5 in]{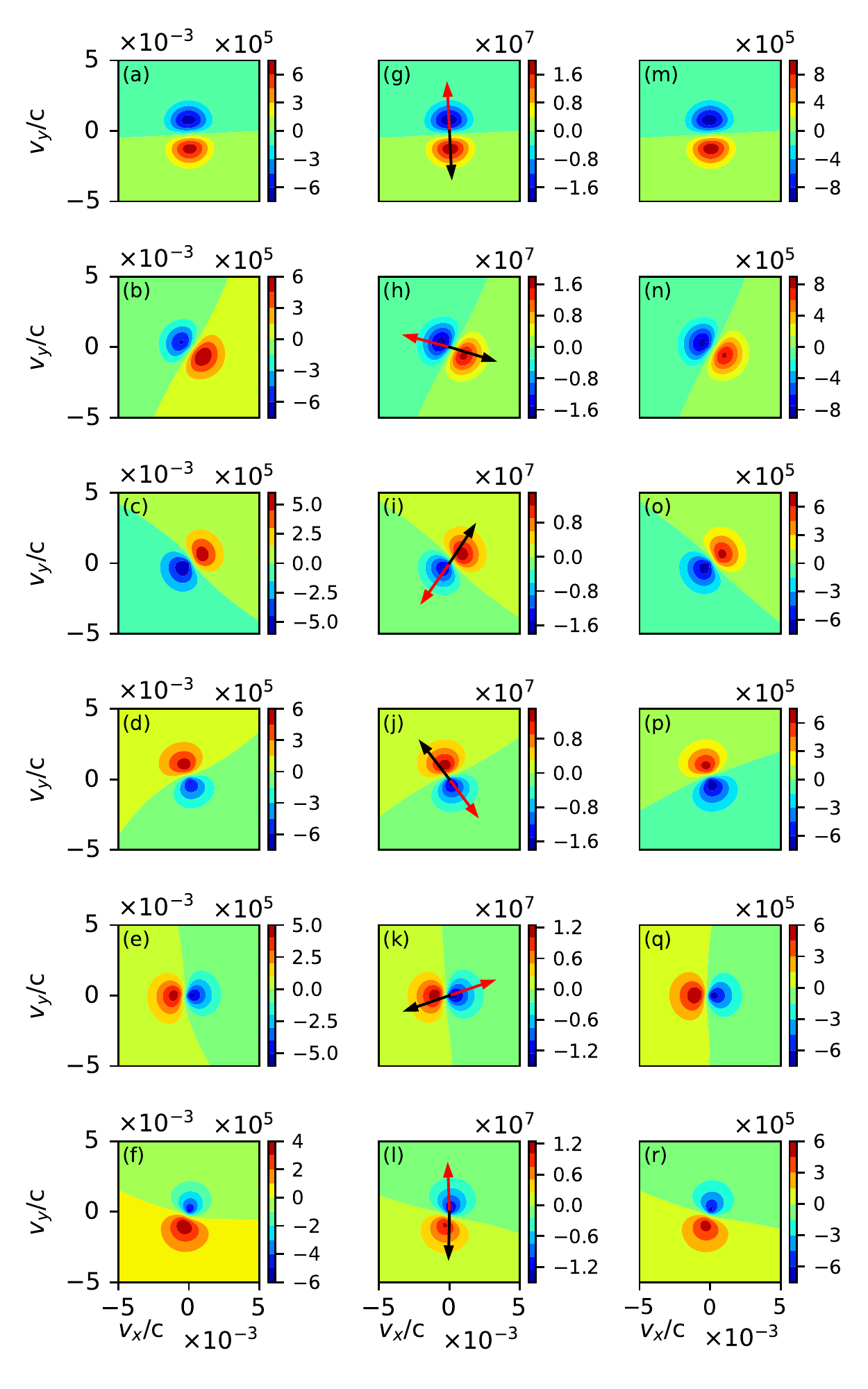}
	\caption{Evolution of the electrons velocity distribution function perturbation in the transverse velocity-space plane $v_{x}$-$v_{y}$ for the cold electrons case, with $v_{th}=0.001 c$. 
	Frames from top to bottom correspond to equal intervals over one whistler cycle ($\sim 20 \omega_{pe}^{-1}$) from $t=0$, while those from left to right are for $v_{z}=-v_{\rm th}$, 0, and $v_{\rm th}$, respectively. 
	Red and black arrows indicate, ${\bf B}_{1}$ and $\langle {\bf v}\rangle_{1}$, respectively.
\label{deltaf1}
}
\end{figure}

\begin{figure}
	\includegraphics[width=3.5 in]{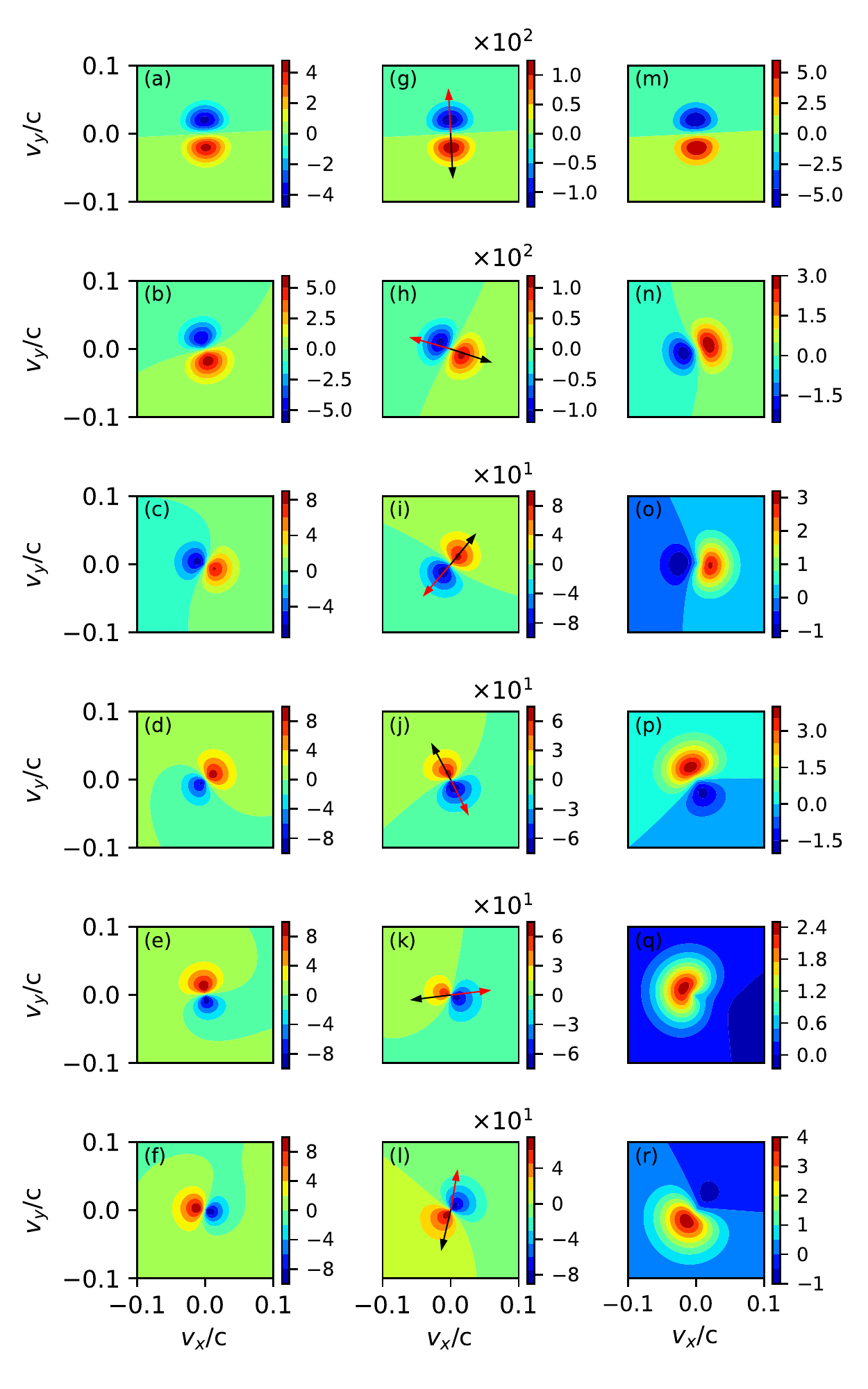}
	\caption{Evolution of the electrons velocity distribution function perturbation in 
	the transverse velocity-space plane $v_{x}$-$v_{y}$ for the warm electrons case, with $v_{th}=0.02 c$. 
	Frames from top to bottom correspond to equal intervals over one whistler cycle ($\sim 20 \omega_{pe}^{-1}$) from $t=0$, while those from left to right are for $v_{z}=v_{\rm res}$, 0, and $v_{\rm phase}$, respectively.
	Red and black arrows indicate, ${\bf B}_{1}$ and $\langle {\bf v}\rangle_{1}$, respectively.
\label{deltaf2}
}
\end{figure}

The plots of electron distribution function perturbation $f_{1}=f-f_{0}$, as 
available from the simulations done with denser grid size,
$64\times 160\times 160 \times 160$, are analyzed in order to understand
the kinetics of the evolution of whistler wave perturbation in the plasma.
Since the system is periodic and transverse, analyzing the $f$ data from 
any chosen spatial location is equivalent. The location $z=L/2$, or the 
center of the simulation domain, is therefore chosen here as the 
representative location in the simulation domain.
For added accuracy, both $f$ and $f_{0}$ are evolved under equivalent 
Vlasov simulation operation in order to obtain $f_{1}(t)$ presented below.
From the full 3-dimensional velocity distribution function available, the 
evolution of $f_{1}(z=L/2,v_{z},v_{x},v_{y})$ at various stages of time
is presented in Fig.~\ref{deltaf1} from the cold electron case simulation, 
performed with sufficiently small $T_{e}$ such that the 
$|v_{\rm res}|\gg v_{\rm th}=0.001~c$. 
Because of the small electron temperature, both the resonant and phase 
velocity of the whistler are far removed from the distribution width 
(along $v_{z}\|{\bf k}$) and
the $f_{1}$ plotted in three columns of frames from left to right 
correspond to velocity values $v_{z}=-v_{\rm th}$, $0$ 
and $v_{\rm th}$, respectively. 
The same data is plotted in Fig.~\ref{deltaf2} from simulation of 
warmer electrons case, $|v_{\rm res}|\le v_{\rm th}=0.001~c$, such that 
sufficient population of resonant electrons is present with 
$v_{z}=v_{\rm res}$, producing dominant and analyzable kinetic effects. 
The rows of frames from top to bottom in both plots correspond to 
time from $t=0$ to 20 $\omega_{pe}^{-1}~(\equiv 2\Omega_{ce}^{-1})$.

In both cold and warm electron cases plotted in top rows of 
Fig.~\ref{deltaf1} and Fig.~\ref{deltaf2}, respectively, 
the $f_{1}$ plotted at $t=0$, consistent with a whistler phase-space 
perturbation in the velocity space, has
a nonzero velocity perturbation 
purely aligned to $\hat{v}_{y}$. 
The following rows cover evolution over a period of
$20\omega_{pe}^{-1}$, or one complete whistler cycle. 
We first discuss evolution of perturbation in the cold electron case 
presented in Fig.~\ref{deltaf1}. Consider the evolution of the perturbation 
component located at $v_{z}\sim 0$ plotted in the second column of the 
Fig.~\ref{deltaf1}.
Because of being drawn from vanishing $v_{z}$, 
this component of $f_{1}$ does not have electrons that leave to, or arrive 
from, the adjacent $z$ locations as the time progresses (i.e., from top frame 
to bottom frame). 
Therefore, the participating electrons are 
simply displaced to newer angular locations with (angular) velocity 
$\Omega_{ce}$ on the same transverse velocity 
plane $v_{x}$-$v_{y}$ in the subsequent frames, with respect 
to their original locations in the $t=0$ frame. 
However the polarization of the perturbation in this case rotates with 
angular velocity of the wave $\omega$, dissimilar to $\Omega_{ce}$ ($>\omega$,
the reason for which is well encoded in the linear theory of waves in 
magnetized plasmas). 
Drawn from the simulation data for ${\bf \langle v\rangle}_{1}$ and ${\bf B}$, 
the black and 
red arrows represent phases of right handed polarized velocity and 
magnetic field vectors, respectively, for the reference.
The rotation in perturbation polarization consistent with wave frequency 
confirms that the initial 
perturbation efficiently excited a whistler eigenmode in the simulated 
magnetized plasma right from $t=0$. 

The kinetic effects because of 
resonant electrons are addressed further below in warm electron case 
since for
cold electrons the velocity distribution drops to negligible values 
at the resonant velocity. 
The same is evident from the identical evolution of the perturbation component 
at $v_{z}=|\pm v_{th}|\ll |v_{\rm res}|$, plotted in the columns 1 and 3, 
respectively, which 
show rotation of perturbation polarization with same frequency and a 
coherent response of electrons of all $v_{z}$ values due to the
distribution function being nearly a delta function in velocities 
both along and across ${\bf k}$. 
Therefore, the fluid theory results ($\gamma\rightarrow 0$, $\omega$ given by (\ref{dispersion-rhcp-cold})) 
remain suitably recoverable for this cold electron case as clear from the 
data point corresponding to $v_{th}\sim 0.001 c$ in Fig.~\ref{comparison}.

We now show that an identical initial perturbation, that excited a 
nearly ideal whistler eigenmode in the above cold electron case, 
evolves in a warm electron magnetized plasma exhibiting a far richer 
kinetic response. 
This initial evolution, covering $20 \omega_{pe}^{-1}$ remains applicable, 
for example, 
to the short lived transverse whistler mode fluctuations characteristically 
running much of weak electromagnetic plasma turbulence \cite{Gary_2010}. 
There evolution therefore necessitates invoking initial phase-mixing
description of analytical kinetic theory results 
in addition to the routine estimates of damping by pure resonant 
electrons. 

Considering the evolution for the perturbation component
at $v_{z}\sim 0, z=L/2$ plotted in the middle column of the frames in 
Fig.~\ref{deltaf2} we note that its polarization 
rotates nearly with the wave frequency $\omega$, however with a 
visible shortfall in comparison to the cold electron case.
The electrons being sufficiently warm in this case, it is now possible to 
examine the $f_{1}$ evolution at the resonant velocity for this set of data,
which is noted to be much distinct as plotted in the first column of the
Fig.~\ref{deltaf2}.

Note that individual electrons streaming with resonant 
velocity $v_{z}$ would see the wave frequency Doppler-shifted to 
$\Omega_{ce}$ and therefore will be perpetually accelerated 
by the wave ${\bf E}_{1}$ in transverse the direction. 
The resulting proliferation of electrons to higher $v_{\perp}$ is visible 
with time in frames from (a)-(f).
It is 
important to notice, though, that, unlike the coherent response of the 
full distribution in cold electron case, the polarization of the perturbation 
in the resonant electrons has rotated with a considerably lower frequency, 
thereby covering a lesser angle as compared to the bulk, or $v_{z}\sim 0$, 
electrons. 

Quantitative estimation of this shift in frequency is possible based on the 
fact that because of $\Omega_{ce}>\omega$, in time $\delta t$ the gyrating 
electrons cover a far larger angle in the transverse plane than the rotating 
polarization of the perturbation. Consequently, as the time progresses
the $v_{z}=v_{\rm res}$ electrons bring imprint of an increasingly older 
phase of the wave, distinct from the instant phase 
of the bulk electron component polarization (seen in the corresponding middle 
column frame plotted for $v_{z}=0$). This is despite of resonant electrons 
arriving (because of their negative streaming velocity $v_{\rm res}<0$) from 
a forward location $z=\delta z>L/2$ having a slightly advanced phase of 
the wave.

The convected phase $\phi(z=L/2,t=\delta t)$ of the polarization for resonant 
perturbation component (perturbation component with $v_{z}=v_{\rm res}$) 
at time $t=\delta t$ can be calculated by using a transformation between 
present and older phase of the polarization of the
perturbation, 
\begin{eqnarray}
	\phi(L/2,\delta t)=\phi(L/2+\delta z, 0)=
	\phi_{0}+\delta \phi_{z}+\delta \phi_{\theta}, 
	\label{transformation}
\end{eqnarray}
where $\phi_{0}$ is instant phase of the polarization of the bulk 
perturbation at $z=L/2$ (as visible in $v_{z}=0$, $t=\delta t$ frame),
$\phi(z,t)$ is the general phase of the polarization 
(ideally known from the cold electron 
limit), 
$\delta \phi_{z}$ and
$\delta \phi_{\theta}$ 
are the phase difference arising from change in $z$ and $\theta$, 
respectively.
For the resonant component of the perturbation, we have,
\begin{eqnarray}
	\delta \phi_{z}=\delta zk = v_{\rm res}(-\delta t)k;~ 
	\delta \phi_{\theta}=(-\delta t)~\Omega_{ce}.
	\label{transformation-z}
\end{eqnarray}
Since $v_{\rm res}<0$ and $\Omega_{ce}>\omega= \Omega_{ce}-|v_{\rm res}|k$,  
we have
$\delta \phi_{z}>0$, $\delta \phi_{\theta}<0$ and 
$|\delta \phi_{\theta}|>\delta \phi_{z}$, meaning that the phase of resonant 
component of the perturbation relative to instant bulk electron phase will be 
negative. These phase delays are in excellent agreement with the phase 
difference in the frames presenting the resonant component of the perturbation 
in the first column of Fig.~\ref{deltaf2}.

An even distinct evolution is witnessed in the perturbation component at the
positive values of parallel velocity, $v_{z}>0$, as potted in third column
of Fig.~\ref{deltaf2} for $v_{z}=v_{\rm phase}$. Before explanation of this
behavior, which strongly relates to phase-mixing process of the whistler 
perturbation, it can be clearly noted that the strength of macroscopic wave 
variables (especially ${\bf J}_{1}$) is going to be weakened with time if 
velocity integration is
performed at advanced times using $f_{1}(v_{x},v_{y})$ that has its peaks 
in $v_{x}$-$v_{y}$ plane initially overlapping for all $v_{z}$, but 
dispersed, at future times, over a wider range of angles for different 
$v_{z}$. 

In the $f_{1}$ evolution plotted at $v_{z}=v_{\rm phase}$ in third column 
of Fig.~\ref{deltaf2}, the streaming (and simultaneously gyrating) electrons 
experience the wave fields purely non-rotating. At each $\delta t$ they 
import, from their original locations 
to $z=L/2$, the phase as determined by the transformation (\ref{transformation}).
In this case, however, with each gyration completed, the electrons 
additionally drift orthogonal to the phase of the wave electric field 
experienced by them.
Consequently, the imported $f_{1}$ is additionally displaced away from 
$v_{\perp}=0$. With time progressing, 
the $f_{1}$ steadily sampled at $z=L/2$ witnesses newer electrons arriving 
from backward locations. 
An increasing drift is therefore
witnessed in $f_{1}$, rotated at an incremented angle in 
$v_{x}$-$v_{y}$ plane. This results in periodically separating red and 
blue patches, absent initially.
Note that the initial transverse velocity perturbation at $t=0$ is 
introduced with no such nonuniformity with respect to $v_{z}$. 
This nonuniformity clearly
develops over the time-scale of an electron gyroperiod. 
\subsection{Kinetic model for initial-value transverse electromagnetic perturbations\label{initial-value-problem}}
Out of the several kinetic modifications of 
$f_{1}$ witnessed above, those attributable 
to the wave fields and those to {\em ballistic} effects can be distinguished
and analyzed based on the linearized Laplace-Fourier transformed 
Vlasov-Maxwell system of equations,
\begin{eqnarray}
	(p&+&i{\bf k \cdot v})f_{kp1}({\bf v})=f_{k1}(t=0)\nonumber\\
	&+&\frac{e}{m_{e}}\left(\frac{\bf v\times B_{0}}{c}\right)\cdot \nabla_{v}f_{kp1}({\bf v})\nonumber\\
	&+&\frac{e}{m_{e}}\left({\bf E}_{kp1}+\frac{{\bf v\times B}_{kp1}}{c}\right)\cdot \nabla_{v}f_{0}({\bf v})
	\label{vlasov-linear}
\end{eqnarray}
\begin{eqnarray}
	{\bf B}_{kp1}={\bf B}_{k1}(t=0)+\frac{c}{ip}({\bf k\times E}_{kp1}).
	\label{maxwell-linear1}
\end{eqnarray}
and
\begin{eqnarray}
	{\bf E}_{kp1}={\bf E}_{k1}(t=0)-\frac{c}{ip}\left({\bf k\times B}_{kp1}-\frac{4\pi}{c}{\bf j}_{kp1}\right),
	\label{maxwell-linear2}
\end{eqnarray}
where $p=-i\omega_{r}+\omega_{i}$ and ${\bf E}_{kp1}$ and ${\bf B}_{kp1}$ are 
the transformed wave electric and magnetic fields, respectively.

Using the well known identity for conjugate components of the velocity 
variable, 
$v_{x}=v_{\perp}\cos{\phi}$ and 
$v_{y}=v_{\perp}\sin{\phi}$,
\begin{eqnarray}\nonumber
	\frac{\partial f_{pk1}({\bf v})}{\partial \phi}
	=\left(
	-v_{y}\frac{\partial}{\partial v_{x}}
	+v_{x}\frac{\partial}{\partial v_{y}}\right)=
	-({\bf v\times \hat{z}})\cdot \nabla_{v} f_{pk1}({\bf v}),
\end{eqnarray}
Eq.~(\ref{vlasov-linear}) can be rewritten as,
\begin{eqnarray}
	\frac{(p+i{\bf k \cdot v})}{\Omega_{ce}}f_{kp1}({\bf v})+
	\frac{\partial f_{pk1}({\bf v})}{\partial \phi}
	=\frac{f_{k1}(t=0)}{\Omega_{ce}}
	\nonumber\\
	+\frac{e}{m_{e}\Omega_{ce}}\left({\bf E}_{kp1}+\frac{{\bf v\times B}_{kp1}}{c}\right)\cdot \nabla_{v}f_{0}({\bf v}).
	\label{vlasov-linear1}
\end{eqnarray}
%
%
\begin{figure}
	\includegraphics[width=3.0 in]{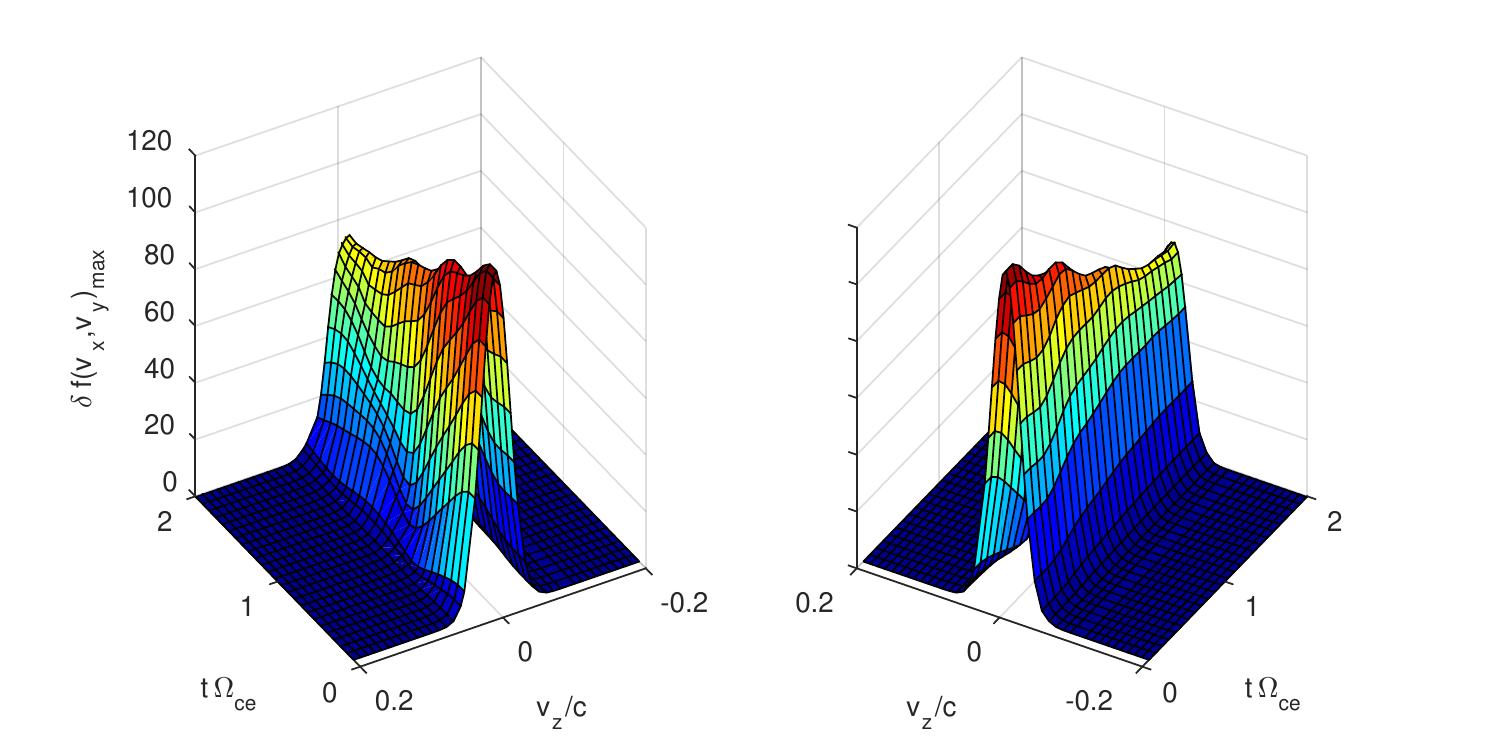}
	\caption{Time evolution of the maximum amplitude, $f_{\rm 1max}$, of electrons velocity distribution function perturbation $f_{1}(v_{x}$,$v_{y})$ plotted in Figs.~\ref{deltaf1} and ~\ref{deltaf2}
in full range of $v_{z}$, for the hot electrons case ($v_{th}=0.02 c$). Plots (a) and (b) present relatively rotated views of the same surface for clearer exposure of the both sides about the surface maximum.
\label{fmax}
}
\end{figure}
\begin{figure}
	\includegraphics[width=3.0 in]{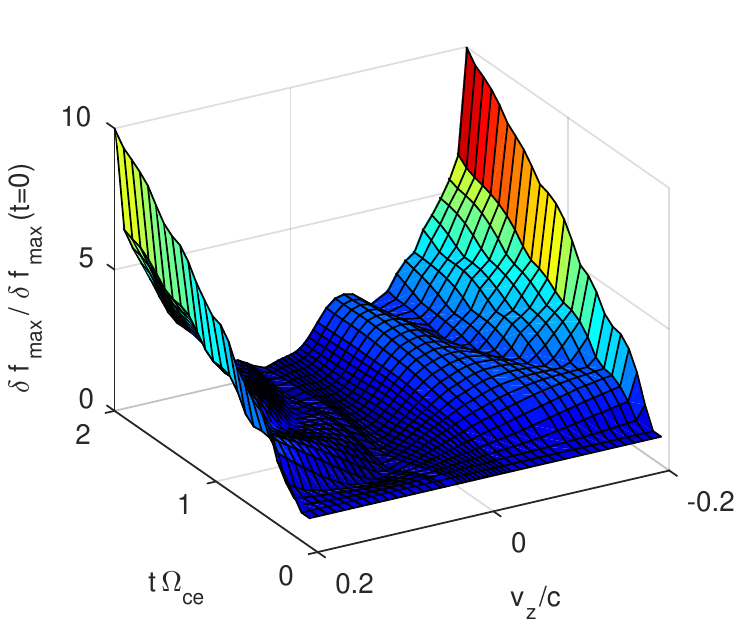}
	\caption{Surface in the figure normalized to individual initial values at each $v_{z}$.
\label{fmaxn}
}
\end{figure}

Note that the wave field generated $v_{z}$ specific modifications, 
including (i) the acceleration of
resonant electrons and (ii) the slowly building electron drift for 
$v_{z}\sim v_{\rm phase}$, can be resolved, approximately, from the 
fundamentally kinetic ballistic modifications in the $f_{1}$ at shorter 
times ($\Omega_{ce}^{-1}<t< \omega^{-1}$) when the latter dominate the former. 
In order to do this, we first write the right handed polarized 
component of the Vlasov equation \cite{bittencourt2013fundamentals} 
applicable to the whistler mode,
\begin{eqnarray}
	\frac{d}{d\phi}\left\{f_{1+}({\bf v})\exp{\left[\frac{(p+ikv_{z})}{\Omega_{ce}}\phi \right]}\right\}=\nonumber\\
	\frac{f_{k1+}(t=0)}{\Omega_{ce}}
	\exp{\left[\frac{(p+ikv_{z})}{\Omega_{ce}}\phi +i\phi\right]}\nonumber\\
	+F_{+}(v_{z},v_{\perp})\exp{\left[\frac{(p+ikv_{z})}{\Omega_{ce}}\phi +i\phi\right]}
	\label{vlasov-rhcp}
\end{eqnarray}
where
$v_{\perp}=(v_{x}^{2}+v_{y}^2)^{\frac{1}{2}}$, $f_{k1+}$ is right handed 
circularly polarized part of the initial perturbation.
The quantity $F_{+}(v_{z},v_{\perp})$ is obtained by ignoring 
${\bf B}_{k}(t=0)$ in Eq.~(\ref{maxwell-linear1}), as, 
\begin{eqnarray}
	F_{+}(v_{z},v_{\perp})=\frac{e}{m_{e}\Omega_{ce}}
	\left[
	\left(1-\frac{kv_{z}}{\omega}\right)
	\frac{\partial f_{0}}{\partial v_{\perp}}
	+\frac{kv_{z}}{\omega}\frac{\partial f_{0}}{\partial v_{\perp}}
	\right]\frac{E_{+}}{\sqrt{2}}
	\nonumber\\
\end{eqnarray}
where $E_{+}=E_{x}-iE_{y}$. Both sides of Eq.~(\ref{vlasov-rhcp}) can be 
integrated with respect to $\phi$ from $\phi=-\infty$ to instant $\phi$
by assuming very small imaginary part in $\omega$, yielding the time 
asymptotic solution,
\begin{eqnarray}
	f_{pk1+}({\bf v})&=&\frac{f_{k1+}(t=0)e^{i\phi}}{(p+ikv_{z}+
	i\Omega_{ce})}\nonumber\\
	&+&\frac{\Omega_{ce}}{(p+ikv_{z}+i\Omega_{ce})}
	F_{+}(v_{z},v_{\perp})e^{i\phi}.
	\label{solution}
\end{eqnarray}
This means that all the right handed circularly polarized transverse modes 
are obtainable by accounting for the poles of the second term, which 
represents the contribution of the electric and magnetic fields of the wave. 
For example, the Whistler mode corresponds to $E_{+}$ obtained in terms of 
$B_{+}$ and $J_{+}$ using (\ref{vlasov-linear})-(\ref{maxwell-linear2}) in 
the limit $\Omega_{pi}<\omega<\Omega_{ce}$ such that the relevant pole appears where 
the dispersion function for an isotropic electron distribution,
\begin{eqnarray}
c^{2}k^{2}-\omega^{2}-\frac{8\pi^{2}q_{e}^{2}}{m}
\int\limits_{-\infty}^{\infty}\int\limits_{0}^{\infty}
\frac{
\left[(\omega-kv_{z}+kv_{\perp})\frac{\partial f_{e0}}{\partial v_{z}}
\right]
}{(\omega-kv_{z}\pm \Omega_{ce})}\nonumber\\
	\times v_{\perp}^{2}dv_{\perp}dv_{z}
	\label{dispersion-function}
\end{eqnarray}
vanishes, yielding both $\omega_{r}$ and $\gamma$ \cite{chen2013improved}.
However, the solution (\ref{solution}) indicates that $f_{pk1+}({\bf v})$,
being the integrand of the 
Landau-like Laplace inversion integral producing 
$f_{k1+}({\bf v},t)$ (which can, in turn, be compared to the simulation 
outputs plotted in Fig.~\ref{deltaf1} and \ref{deltaf2}), 
has an additional pole at,
\begin{eqnarray}
	p=-i(kv_{z}+\Omega_{ce}). 
	\label{pole}
\end{eqnarray}
Since $k$, $v_{z}$ and $\Omega_{ce}$ are real, there is no 
reduction expected with time in the amplitude of perturbation components 
corresponding to 
various $v_{z}$. Instead, $f_{k1+}(v_{z},t)$ must develop temporal 
oscillations with a pure real frequency $\omega_{b}(v_{z})$
directly proportional to $v_{z}$. As seen below, this temporal modulation
is caused by the information of the wave phases at remote locations being 
brought by the electrons
at the location of measurement,
at different temporal rates for different $v_{z}$. 

The net (kinetic) reduction in the wave strength is therefore caused by 
both, (i) a
phase-mixing process present throughout the velocity space and (ii) the
resonant damping which is dominant at the resonant velocity,
accelerating only the resonant electrons.
While both these processes in a collisionless warm plasma conserve
the entropy despite diminishing the wave strength \cite{krall:na}, 
only the latter contributes to the routinely estimated $\gamma$.
The temporal damping caused by
former involves pure frequencies and cause a rather {\em deterministic
thermalization} (reversible or entropy preserving ergodization), 
or a {\em phase-mixing}, of the transverse distribution (and, consequently of
${\bf J}_{+}$). As presented below, this phase-mixing is clearly recoverable 
from the output of a deterministic simulation performed in this study, duly 
described by the above analytical treatment.
%
\begin{figure}
	\includegraphics[width=2.0 in]{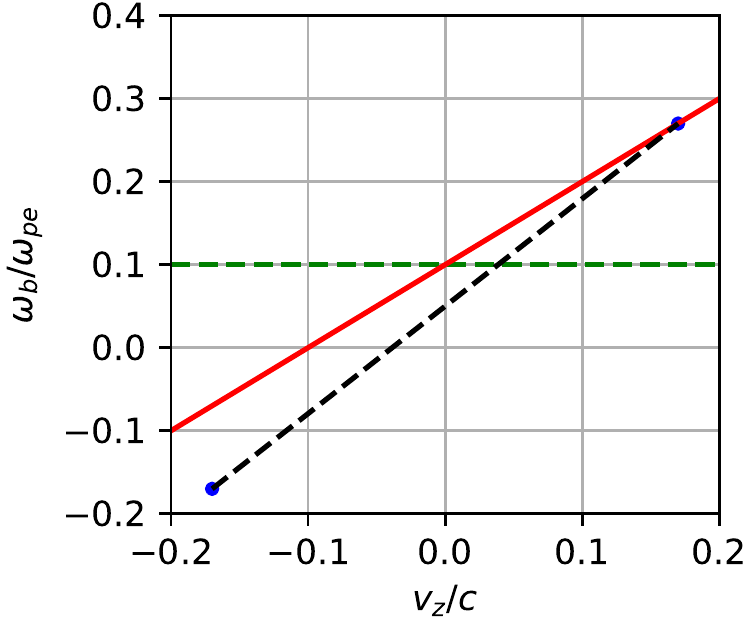}
	\caption{The simulated $v_{z}$ dependence frequency $\omega_{b}$ (dotted line, circular markers) estimated from the Fig.~\ref{fmaxn} compared with analytical expression (\ref{pole}) (solid line) for frequency of amplitude oscillation resulting from phase-mixing. 
\label{wb}
}
\end{figure}

The maxima of $f_{1}(v_{x},v_{y})$ presented in Fig.~\ref{deltaf2}, 
corresponding to hot electron case, are plotted separately as a function of 
$v_{z}$ and $t$ 
in the form of the surface $f_{\rm 1max}(v_{z},t)$ in Fig.~\ref{fmax}. For the 
convenience, two views of the surface are presented in (a) and (b)
with relative rotation to clearly highlight the visible temporal variation 
of $f_{\rm max}$ at various $v_{z}$, including $v_{z}=v_{\rm res}$ as 
indicated by an arrow.
The perturbation amplitude in Fig.~\ref{fmax} shows that only at 
$v_{z}=v_{\rm res}$ a nearly monotonic variation is present in the
amplitude, whereas at velocities away from $v_{\rm res}$ a slow
temporal oscillation enters in the amplitude of $f_{1}$ with its frequency 
increasing with $|v_{z}|$. The high frequency oscillations in small $v_{z}$ 
region are remnants of high frequency ($\omega_{R,L}>\Omega_{ce}$) RHCP 
and LHCP waves (see Fig.~\ref{decay1}) and should not be confused with 
the discussed low 
frequency high amplitude oscillations in the small $v_{z}$ region. 

This additional frequency in the surface plot contours for larger $v_{z}$ 
$(>0)$ results from the streaming electrons importing information of 
polarization from adjacent locations based on the transformation 
Eq.~(\ref{transformation}). Note that because of the periodic set up, for 
a larger velocity, say $v_{z}\sim n(k\delta t)^{-1}$, the information being 
imported would contain $n$ cycles of modulation, 
corresponding to $n$ consecutive spatial periods sampled by fast streaming 
electrons in the time interval $\delta t$ past $t=0$.
The frequency of temporal modulation in 
$f_{\rm 1max}(v_{z},t)$ therefore must increase with $v_{z}$ value being 
examined at a fixed location $z=L/2$. 
This dependence on $v_{z}$ is indeed confirmed by plotting, in 
Fig.~\ref{fmaxn}, the amplitude $f_{\rm 1max}(v_{z},t)$ 
normalized to its $t=0$ reference value for each $v_{z}$, such that the 
modulations become clearly identifiable even at the larger $v_{z}$
where the distribution function $f_{0}$ as well as the amplitude of 
$f_{1}$ drop to significantly small values.

The perturbation information at $z=L/2$ arriving loaded on electrons streaming 
with phase velocity is played faster than the rotation of bulk polarization 
(as $\omega<\Omega_{ce}$) but imports an increasingly delayed phase (causing 
the rotation appearing to stop or reversed over certain intervals). 
Additionally, after certain time interval, it begins to display the 
signatures of freshly developed ${\bf E}_{1}\times {\bf B}_{0}$ drift. 

Finally, since electrons with $v_{z}> v_{\rm res}$ are effectively magnetized
to a higher degree on the wave time scale experienced by them, 
$(\omega-kv_{z})^{-1}\gg \Omega_{ce}^{-1}$, 
they participate rather efficiently in the collective wave propagation.
However, a Doppler down-shifted frequency experienced, and responded to,
by this majority of electrons causes
the wave collective $\omega_{r}$ to reduce in
accordance with dispersion (\ref{dispersion-rhcp}). This reduction in 
$\omega_{r}$ is duly 
captured by the present Vlasov simulation results, as clear by copmparing 
Fig.~\ref{deltaf1} with Fig.~\ref{deltaf2}.
%
\section{Summary and Conclusions \label{conclusion}}
The kinetic damping mechanism of low frequency transverse perturbation propagating parallel to the magnetic field in a magnetized warm plasma is simulated by means of Vlasov simulations. The impact of finite temperature on the wave frequency and the underlying physics of the analytical approximations prescribing reduction in the real frequency of the wave in a warm plasma is addressed and resolved by full 4D phase-spatiotemporal evolution simulated of the electron distribution function.  

The analysis of the Vlasov simulation output illustrates and estimates 
the evolution of initially coherent perturbations that undergo damping 
by mutually separable initial kinetic evolution operating over the global and 
local regions of the of electron velocity space.
The net kinetic effects contributed reduction in the strength of perturbations
is caused by both, (i) a phase-mixing process present throughout the velocity 
space and (ii) the resonant damping which is dominant at the resonant 
velocity, accelerating only the resonant electrons. 
The quantitative analysis of the amplitude oscillation produced by the former 
is done and their interplay, at short times, with the modifications of the 
initial perturbations by the resonant electrons is shown to result in enhanced 
decay of the perturbation amplitude as compared to typical long time estimates
of the damping. It is emphasized that while both these mechanisms in a 
collisionless warm plasma conserve the entropy, it is only the resonant
effects that contributes to the routinely estimated damping rate $\gamma$.
The analytically predicted reduction in the real wave frequency $\omega_{r}$
is duly recovered by the simulations. This is shown to follow
from a greater effective magnetization of dominant $v_{z}> v_{\rm res}$ 
electron population, which, however, experiences frequency of the wave 
Doppler-shifted to smaller values. 

We conclude by discussing that access to rather self-consistent 
analytical initial value solutions of the problem
(\ref{vlasov-linear})-(\ref{maxwell-linear2}) remain of high relevance. 
They should enable
the output from the present fully kinetic electromagnetic transverse wave 
Vlasov simulations to more carefully determine, for example, the spectral 
boundaries of the electromagnetic collisionless plasma turbulence. 
The associated particle and energy transport in the space and fusion 
plasma, in turn, can also be approached more deterministically.

\noindent
Acknowlegement: The simulations are performed on the IPR HPC supercomputing 
cluster ANTYA.
\bibliography{vlasov.bib}
\end{document}